\documentclass[twocolumn,prx,showpacs,superscriptaddress,preprintnumbers,amsmath,amssymbl]{revtex4-1}

\usepackage{amssymb}
\usepackage{graphicx}
\usepackage{dcolumn}
\usepackage{bm}
\usepackage{color}

\begin{document}


\title{A New Ferromagnetic Superconductor: CsEuFe$_4$As$_4$}

\author{Yi Liu} \affiliation{Department of Physics, Zhejiang University, Hangzhou
310027, China} \affiliation{Collaborative Innovation Centre of Advanced Microstructures, Nanjing 210093, China}

\author{Ya-Bin Liu} \affiliation{Department of Physics, Zhejiang University, Hangzhou
310027, China} \affiliation{Collaborative Innovation Centre of Advanced Microstructures, Nanjing 210093, China}

\author{Qian Chen} \affiliation{Department of Physics, Zhejiang University, Hangzhou
310027, China} \affiliation{Collaborative Innovation Centre of Advanced Microstructures, Nanjing 210093, China}

\author{Zhang-Tu Tang} \affiliation{Department of Physics, Zhejiang University, Hangzhou
310027, China} \affiliation{Collaborative Innovation Centre of Advanced Microstructures, Nanjing 210093, China}

\author{Wen-He Jiao} \affiliation{Department of Physics, Zhejiang University of Science and Technology, Hangzhou 310023, China}

\author{Qian Tao} \affiliation{Department of Physics, Zhejiang University, Hangzhou
310027, China} \affiliation{Collaborative Innovation Centre of Advanced Microstructures, Nanjing 210093, China}

\author{Zhu-An Xu}
\affiliation{Department of Physics, Zhejiang University, Hangzhou
310027, China} \affiliation{Collaborative Innovation Centre of Advanced Microstructures, Nanjing 210093, China}

\author{Guang-Han Cao} \email[]{ghcao@zju.edu.cn}
\affiliation{Department of Physics, Zhejiang University, Hangzhou
310027, China} \affiliation{Collaborative Innovation Centre of Advanced Microstructures, Nanjing 210093, China}

\date{\today}

\begin{abstract}
Superconductivity (SC) and ferromagnetism (FM) are in general antagonistic, which makes their coexistence very rare. Following our recent discovery of robust coexistence of SC and FM in RbEuFe$_4$As$_4$ [Y. Liu et al., arXiv: 1605.04396 (2016)], here we report another example of such a coexistence in its sister compound CsEuFe$_4$As$_4$, synthesized for the first time. The new material exhibits bulk SC at 35.2 K and Eu$^{2+}$-spin ferromagnetic ordering at 15.5 K, demonstrating that it is a new robust ferromagnetic superconductor.

\end{abstract}
\maketitle
\section{Introduction}

The search for ferromagnetic superconductors (FMSCs) can be trace back to before the 1960s\cite{ginzburg}. Owing to the antagonistic nature of superconductivity (SC) and ferromagnetism (FM)\cite{ginzburg,buzdin85}, SC rarely coexists with FM, even for that SC and FM emerge in different subsystems of a complex crystalline lattice. It was not until late 1970s that both SC and FM were observed in ErRh$_4$B$_4$, but SC \emph{disappears} when the Er magnetic ordering sets in\cite{ErRh4B4}. Since then, a few `conventional magnetic superconductors' were discovered\cite{maple}, in which SC and local-moment FM (or more frequently, other types of magnetic orderings with ferromagnetic components) casually coexist in certain temperature and magnetic regimes. Such materials were earlier called FMSCs\cite{buzdin85}, and this terminology was also employed for the uranium compounds UGe$_2$, URuGe and UCoGe that are superconducting well below their Curie temperatures\cite{huxley}. Here we adopt the classification\cite{chu}, which gives another terminology, `superconducting ferromagnet', for the case that the superconducting transition temperature $T_{\mathrm{sc}}$ is lower than the Curie temperature $T_{\mathrm{Curie}}$. Whilst FMSC is reserved for the scenario of $T_{\mathrm{sc}}>T_{\mathrm{Curie}}$. Note that, in the U-based superconducting ferromagnets, SC and FM share the same type of electrons, and the SC is widely believed to be in a spin-triplet state. For a spin-singlet superconductor, however, SC is more easily destroyed by the strong exchange fields in a ferromagnet. At the same time, a spin-singlet $s$-wave superconducting state does not allow a local-moment FM via Ruderman-Kittel-Kasuya-Yosida (RKKY) interactions\cite{anderson}. Therefore, it seems impossible for a single material to host both local-moment FM and spin-singlet SC (FM+SC).

However, the FM+SC-like phenomenon was observed\cite{ren2009,cao2011}, and recently confirmed by x-ray resonant magnetic scattering\cite{rxs2014} and neutron scattering\cite{nd2014}, in the P-doped EuFe$_2$As$_2$ system in which SC emerges at $\sim$26 K followed by ferromagnetic ordering at $\sim$17 K for the Eu$^{2+}$ spins. Similar phenomena were later demonstrated in other doped EuFe$_2$As$_2$ systems\cite{jiang2009,jin-Co,jiao2011,jiao2013,jin-Ir}. Nevertheless, strong experimental evidence of bulk SC in the  doped EuFe$_2$As$_2$ systems is still lacking. Also there have been debates on the nature of Eu-spin ordering\cite{cao2011,rxs2014,nd2014,felner2011,jeevan2011,zapf2013}.

Very recently, motivated by our previous material design\cite{jh} as well as the latest experimental progress\cite{1144}, we succeeded in synthesizing a new Eu-containing iron arsenide RbEuFe$_4$As$_4$ which exhibits bulk SC at $T_{\mathrm{sc}}$ = 36.5 K and Eu-spin FM at $T_{\mathrm{Curie}}$ = 15 K\cite{Rb1144}. The robustness of both SC and FM indicates a genuine FM+SC state realized. Here we report the second robust FMSC, CsEuFe$_4$As$_4$, a sister compound of RbEuFe$_4$As$_4$. The new material shows bulk SC at 35.2 K and Eu-spin FM at 15.5 K. An additional anomaly at 5 K is observed, possibly associated with the interplay between SC and FM. Another interesting issue is that the Eu-spin ferromagnetic ordering is of a rare third order, suggesting a strong two-dimensional character of the ferromagnetic transition.

\section{Experimental}
CsEuFe$_4$As$_4$ polycrystalline sample was synthesized by solid-state reactions in a sealed vacuum, with procedures similar to the synthesis of RbEuFe$_4$As$_4$ \cite{Rb1144}. First, CsAs, EuAs and FeAs were prepared respectively via the reactions of Cs (99.75\%), Eu (99.9\%) and As (99.999\%) pieces with Fe powders (99.999\%). The intermediate products were then ball milled separately for 10 minutes in a glove box filled with pure Ar (the water and oxygen content is below 1 ppm). Second, the powder of CsAs, EuAs, FeAs and Fe was weighed in a stoichiometric ratio. The mixture was homogenized by grinding, pressed into pellet, and then loaded in an alumina tube which was sealed by arc welding in argon atmosphere in a Ta tube. The welded Ta tube was jacketed in a quartz ampoule filled with Ar gas ($\sim$0.6 bar), followed by heating the ampoule to 1123-1173 K, holding for 6 hours, in a muffle furnace. The sample was quenched after the high-temperature chemical reactions.

Powder x-ray diffraction (XRD) was carried out at room temperature on a PANalytical x-ray diffractometer
(Model EMPYREAN) with a monochromatic CuK$_{\alpha1}$ radiation. To avoid severe preferred orientations, the powder was pressed softly on the sample holder. The lattice parameters and the atomic positions were refined by a Rietveld analysis using a RIETAN-FP software \cite{Rietan-fp2}. The electrical and heat-capacity measurements were conducted on a physical property measurement system (PPMS-9, Quantum Design). The electrical resistivity was measured using a standard four-electrode method. The as-prepared CsEuFe$_4$As$_4$ pellet was cut into a thin rectangular bar, on which gold wires were attached with silver paint. The Hall coefficient was measured by permutating the
voltage and current electrodes under 8 T\cite{Hall}, using a thin-square sample (2.2$\times$2.0$\times$0.17 mm$^3$) with four symmetric electrodes attached. The heat capacity was measured by a thermal relaxation method using a square-shaped sample plate with a total mass of 16.8 mg. The dc magnetization was measured in a magnetic property measurement system (MPMS-5, Quantum Design) using a regular shape sample so that the demagnetization factor can be estimated more precisely.

\section{Results and discussion}
\subsection{Crystal structure}

The powder XRD pattern of the CsEuFe$_4$As$_4$ sample can be well indexed with a RbCaFe$_4$As$_4$-type\cite{1144} (1144-type) primitive tetragonal lattice. No evident impurity phase can be identified. Fig. 1 shows the Rietveld refinement profile based on the 1144-type structure shown in the inset.  The refinement yields a weighted reliable factor $R_{\mathrm{wp}}$ of 4.80$\%$ and a "goodness-of-fit" parameter $S$ of 1.48, and the resulting crystallographic parameters are tabulated in Table~\ref{structure}.
\begin{figure}
\includegraphics[width=8cm]{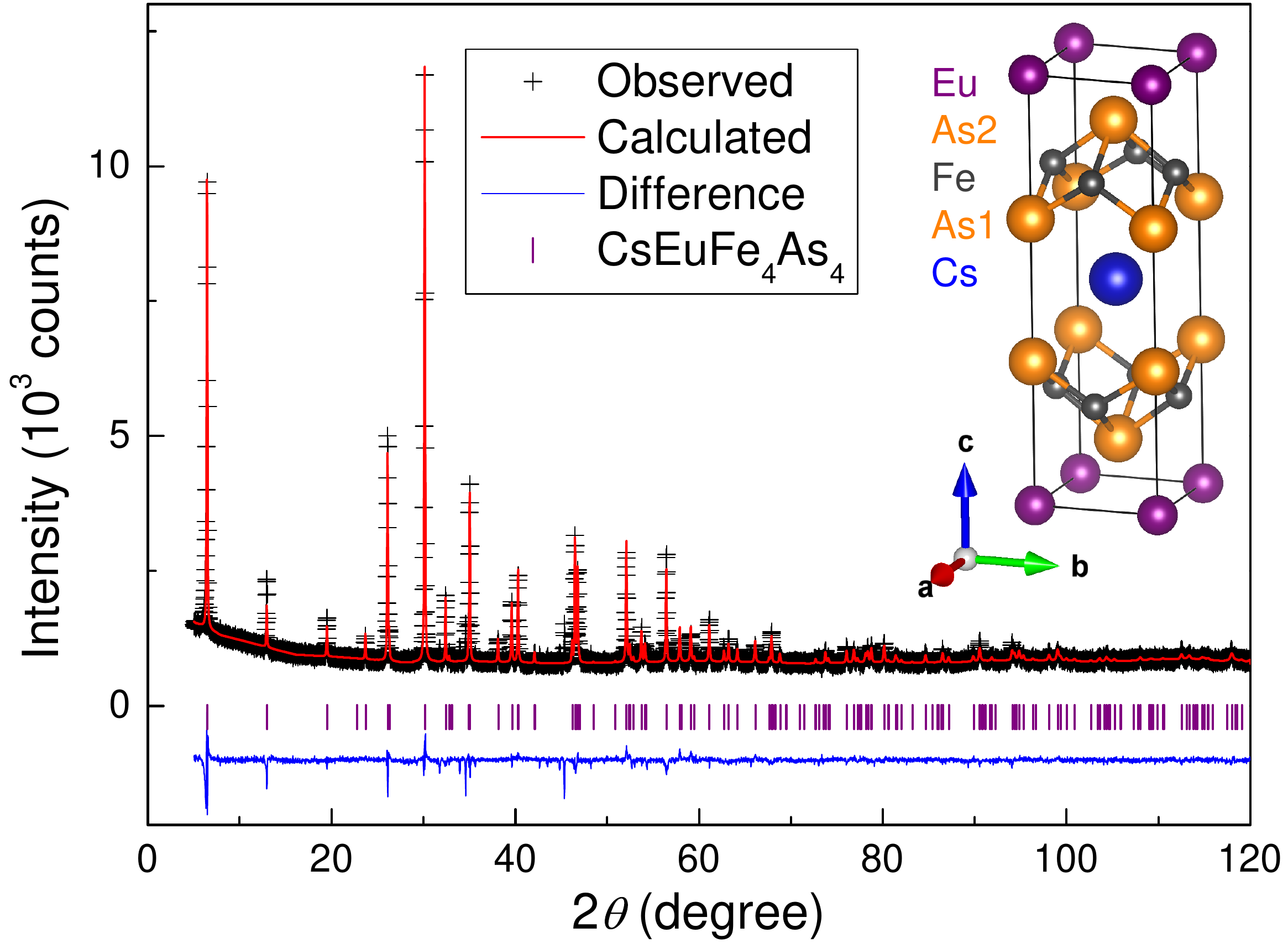}
\label{fig1} \caption{Powder X-ray diffraction and its Rietveld refinement profile for CsEuFe$_4$As$_4$. The inset shows the crystal structure.}
\end{figure}
Isostructural to RbEuFe$_4$As$_4$, CsEuFe$_4$As$_4$ can be viewed as an intergrowth of CsFe$_2$As$_2$ and EuFe$_2$As$_2$, thus it is meaningful to compare their crystal structures. The $a$ axis is almost the same (within the determination uncertainty) as the average value of those of EuFe$_2$As$_2$\cite{ren2008} and CsFe$_2$As$_2$\cite{Cs122}. In fact, the lattice mismatch between the two 122-type compounds is only 0.4\%, which possibly plays an important role for the formation of CsEuFe$_4$As$_4$\cite{jh,1144}. However, the $c$ axis is 0.010(1) \r{A} larger that the sum of one half of the $c$ axes of CsFe$_2$As$_2$ and EuFe$_2$As$_2$. This result looks abnormal because in general the $c$ axis of the hybrid phase should be shortened (as indeed seen in RbEuFe$_4$As$_4$\cite{Rb1144}), in order to stabilize the intergrowth compound. One may examine the lattice changes in each crystalline ``block". The thickness of the ``CsFe$_2$As$_2$" block is almost identical, while the "EuFe$_2$As$_2$" block is slightly stretched from 6.068 \r{A} to 6.082 \r{A}. This small change contrasts with the case in RbEuFe$_4$As$_4$ where the ``RbFe$_2$As$_2$" block is appreciably compressed\cite{Rb1144}.

\begin{table}
\caption{Crystallographic data of CsEuFe$_4$As$_4$ at room temperature with $a$ = 3.9002(1) \r{A}, $c$ = 13.6285(4) \r{A} and space group $P4/mmm$ (No. 123).}
\label{structure}       
\begin{tabular}{cccccc}
\hline\noalign{\smallskip}
Atom& Wyckoff& $x$ &$y$&$z$&$B (\mathrm{\r{A}^{-2}})$ \\
\noalign{\smallskip}\hline\noalign{\smallskip}
Eu & 1$a$&0& 0  &0  &1.3(1)\\
Cs & 1$d$&0.5& 0.5 &0.5  &1.6(1) \\
Fe & 4$i$&0&0.5  &0.2231(2) &0.4(1)\\
As1 &2$g$&0&  0 &0.3229(3) &1.7(1)\\
As2 &2$h$&0.5&0.5   &0.1238(2) &1.5(1)\\
\noalign{\smallskip}\hline
\end{tabular}
\end{table}

The $a$ and $c$ axes of CsEuFe$_4$As$_4$ are 0.22\% and 2.26\% larger, respectively, than the counterparts of RbEuFe$_4$As$_4$, which is obviously due to the incorporation of a larger Cs$^+$ cation. This means that some sort of negative chemical pressures exist in CsEuFe$_4$As$_4$, or equivalently speaking, positive chemical pressures are present in RbEuFe$_4$As$_4$. Note that the Fe coordination in RbEuFe$_4$As$_4$ is asymmetric, characterized by obviously unequal Fe$-$As1 and Fe$-$As2 bondlengths. Here for CsEuFe$_4$As$_4$, however, the Fe$-$As1 and Fe$-$As2 bondlengths are almost equal. Consequently, As1 (close to Cs) and As2 (nearby Eu) heights from the Fe plane are both 1.36(1) {\AA}, and the bond angles As1$-$Fe$-$As1 and As2$-$Fe$-$As2 are 110.3(2)$^{\circ}$. In comparison, the As$-$Fe$-$As angles of CsFe$_2$As$_2$ and EuFe$_2$As$_2$ are 109.6$^{\circ}$ and 106.8$^{\circ}$, respectively. This indicates that, although there is no lattice shrinkage in CsEuFe$_4$As$_4$, the local structure of Fe$_2$As$_2$ layers is ``homogenized", which may account for the lattice stabilization.

\subsection{Transport properties}

Figure 2(a) shows the temperature dependence of electrical resistivity [$\rho(T)$] and Hall coefficient [$R_\mathrm{H}(T)$] for the CsEuFe$_4$As$_4$ polycrystalline sample. The normal-state $\rho(T)$ curve displays an unusual metallic behavior characterized by a broad hump at $\sim$ 150 K. The broad hump seems to be a common feature of hole-doped iron-based superconductors\cite{Ba122K,Sr122K,Ba122K-SB,wutao}. Indeed, the normal-state Hall coefficient $R_\mathrm{H}(T)$ is positive with a value of 6 $\times$ 10$^{-4}$ cm$^3$ C$^{-1}$ at room temperature. If assuming a single-band model, the $R_\mathrm{H}$ value corresponds to 0.54 holes/Fe, more than twice of the expected value (0.25) from charge balance. This discrepancy probably comes from the electron-hole compensation effect because there are multiple bands in the system\cite{Ba122K-SB}. Note that $R_\mathrm{H}(T)$ also shows a hump at around 150 K, suggesting a crossover in electronic properties. This phenomenon is similar to those in heavily hole-doped $A$Fe$_2$As$_2$ ($A$ = K, Rb and Cs) which is revealed as an emergent Kondo lattice behavior very recently\cite{wutao}.

\begin{figure}
\includegraphics[width=8cm]{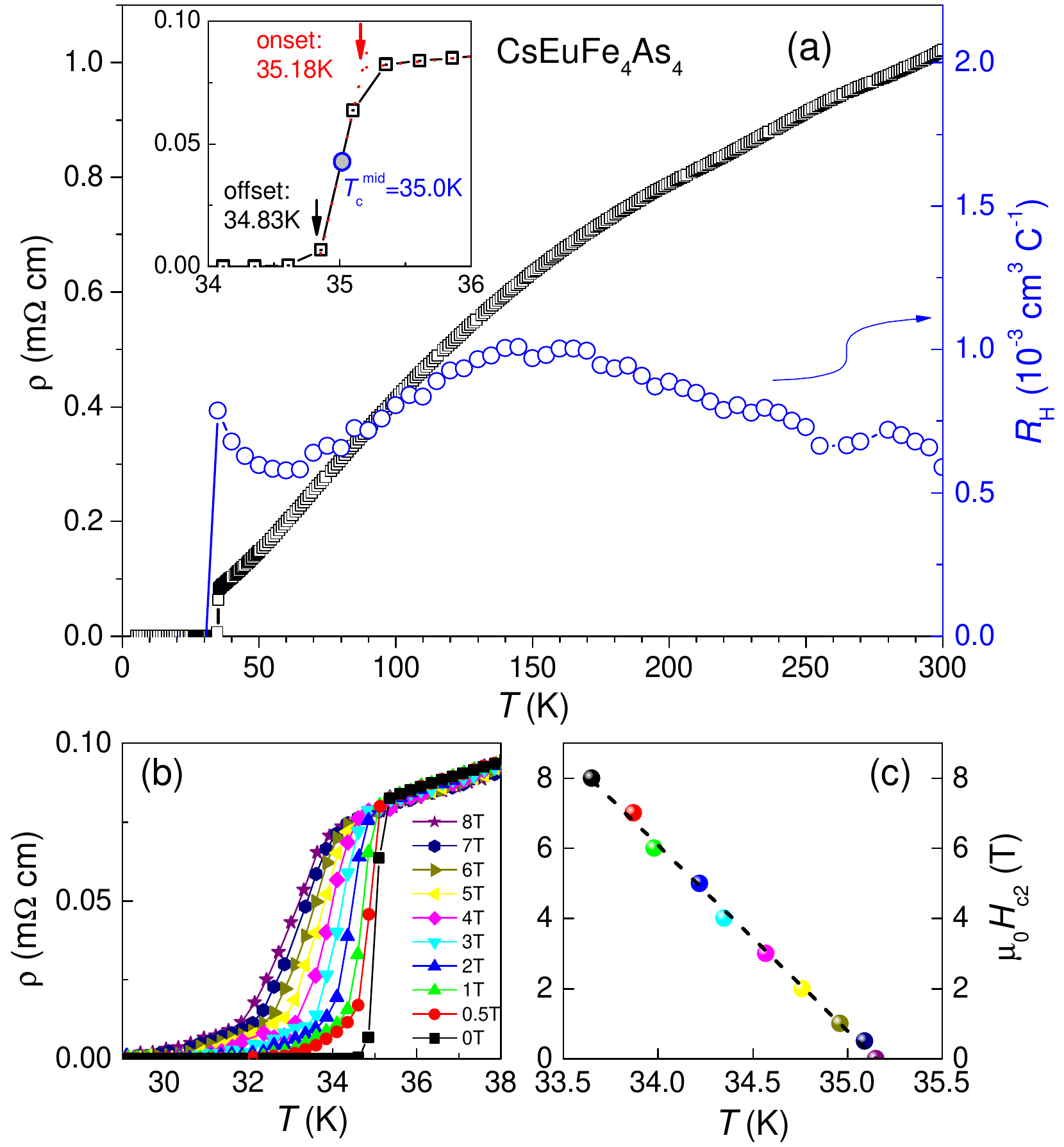}
\label{fig2} \caption{(a) Electrical resistivity (left axis) and Hall coefficient (right axis) as functions of temperature for the CsEuFe$_4$As$_4$ polycrystals. (b) Superconducting transitions under external magnetic fields. (c) Temperature dependence of the extracted upper critical field.}
\end{figure}

As can be clearly seen in the inset of Fig. 2(a), a superconducting transition occurs at $T_{\mathrm{sc}}^{\mathrm{onset}}$ = 35.2 K or $T_{\mathrm{sc}}^{\mathrm{mid}}$ = 35.0 K. Below $T_{\mathrm{sc}}$, it exhibits a zero-resistance state as usual, in contrast with the re-entrant phenomenon observed in most doped EuFe$_2$As$_2$ systems\cite{ren2009,jiao2011,jeevan2011}. Upon applying magnetic fields, $T_{\mathrm{sc}}$ decreases gradually [Fig. 2(b)]. The upper critical fields, $H_{\mathrm{c2}}(T)$, are obtained by defining $T_{\mathrm{sc}}(H)$ as the temperature at which the resistivity drops to 90\% of the extrapolated normal-state value. The extracted $H_{\mathrm{c2}}(T)$ is plotted in Fig. 2(c), which shows a linear relation with an initial slope of $\mu_0$d$H_{\mathrm{c2}}$/d$T$ = $-$5.3 T/K. This $H_{\mathrm{c2}}(T)$ slope is close to that in Sr$_{0.6}$K$_{0.4}$Fe$_2$As$_2$\cite{wnl}, but nearly four times of that of EuFe$_2$(As$_{0.7}$P$_{0.3}$)$_2$\cite{ren2009}, suggesting that SC in CsEuFe$_4$As$_4$ is hardly influenced by the Eu-spin magnetism.

\subsection{Magnetic properties}

Figure 3 shows the temperature dependence of dc magnetic susceptibility [$\chi(T)$] in ZFC and FC modes under $H$ = 10 Oe. A sharp diamagnetic transition occurs at $T_{\mathrm{sc}}^{\mathrm{onset}}$ = 35.2 K, consistent with the above electrical measurement. The volume fraction of magnetic shielding achieves 95\% at 2 K, after the correction of demagnetization effect, and the volume fraction of magnetic repulsion reaches 23\% at $\sim$25 K, indicating bulk SC in CsEuFe$_4$As$_4$.

\begin{figure}
\includegraphics[width=8cm]{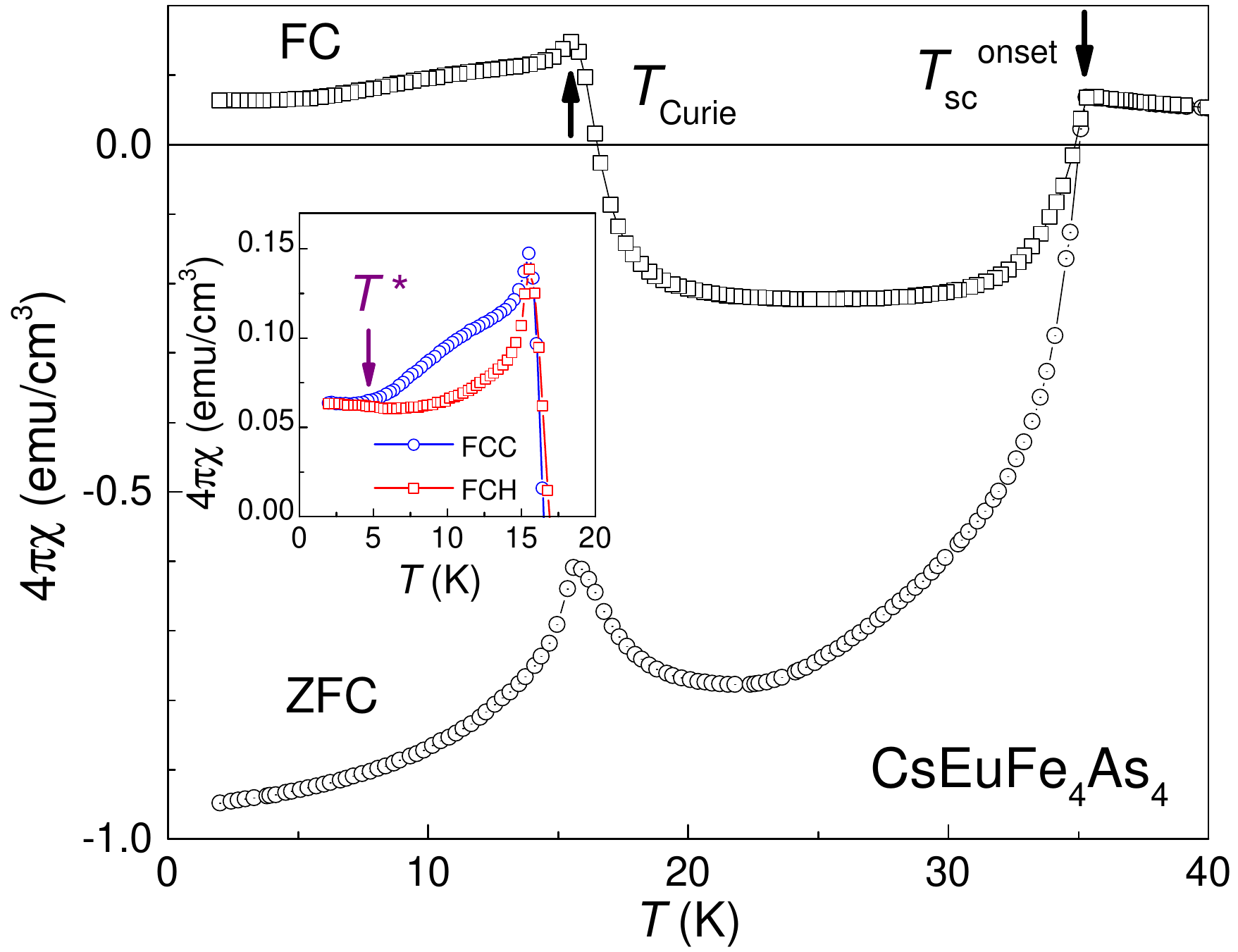}
\label{fig3} \caption{Temperature dependence of magnetic susceptibility for CsEuFe$_4$As$_4$ measured in field-cooling (FC) and zero-field-cooling (ZFC) modes. Superconducting transition at $T_{\mathrm{sc}}^{\mathrm{onset}}$ = 35.2 K and magnetic transition at $T_{\mathrm{Curie}}$ = 15.5 K are marked by arrows, respectively. The inset shows a bifurcation in the FC data measured in cooling (FCC) and heating (FCH) processes. The bifurcation temperature is at $T^*$ = 5 K.}
\end{figure}

Below 18 K, $\chi_{\mathrm{FC}}$ increases abruptly down to 15.5 K where it forms a peak. At the same time, $\chi_{\mathrm{ZFC}}$ also shows a peak at the same temperature. The result is very similar to that in RbEuFe$_4$As$_4$ \cite{Rb1144}. The strong magnetic signals in the superconducting state come from the Eu-spin ferromagnetic ordering, as evidenced by the isothermal magnetization [$M(H)$] shown in Fig. 4. The upturn in $\chi$ above $T_{\mathrm{Curie}}$ = 15.5 K seems to be due to strong two-dimensional correlations (see the specific-heat result below), while the drop in $\chi$ below $T_{\mathrm{Curie}}$ could be due to an antiferromagnetic coupling between the ferromagnetic domains. Additionally, we find a novel bifurcation of FCC (measured in cooling process) and FCH (measured in heating process). However, the FCC and FCH curves merge together below $T^*$ = 5 K. The observation could reflect the interplay between SC and FM. We will discuss the possible origin later.

\begin{figure}
\includegraphics[width=8cm]{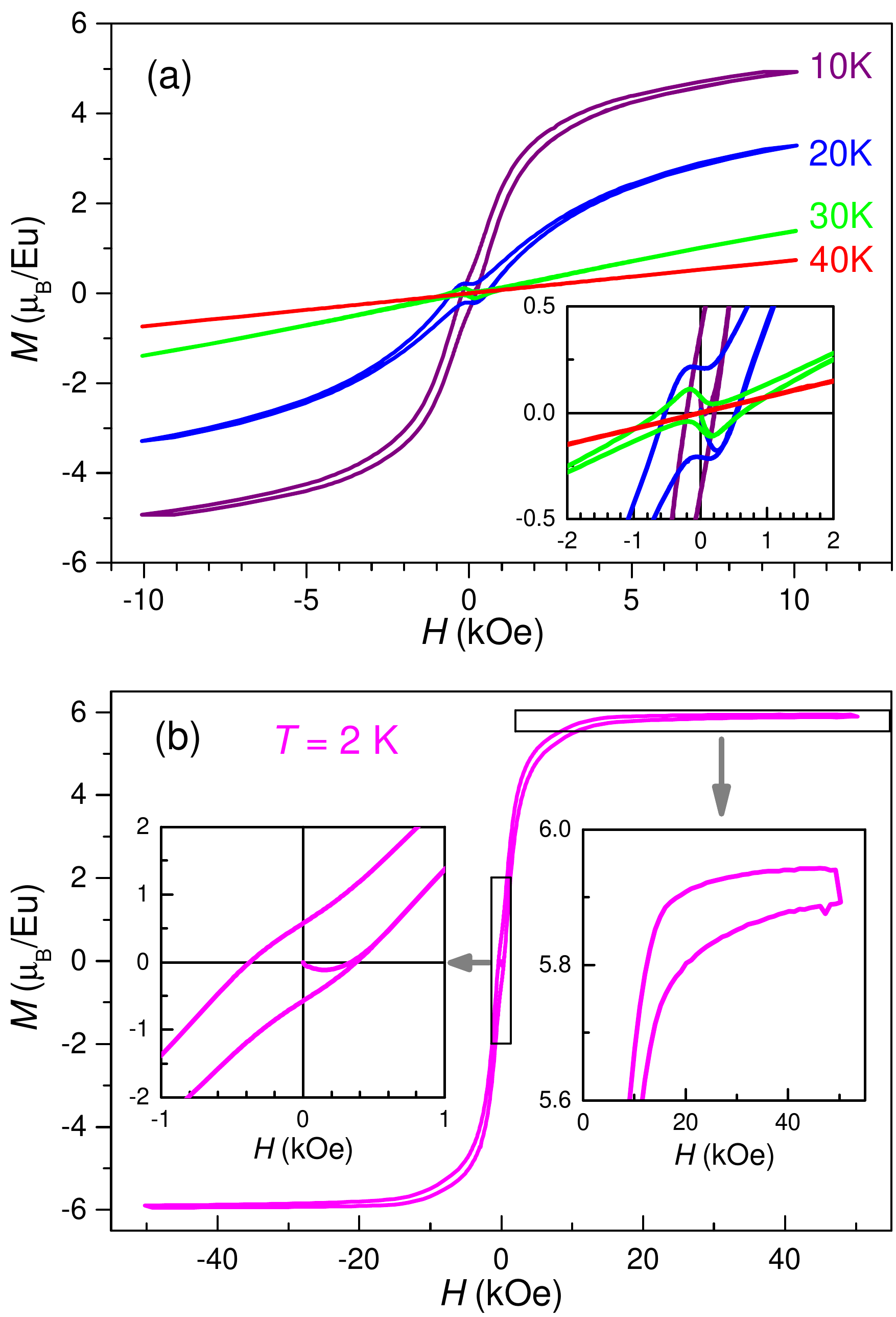}
\label{fig4} \caption{Isothermal magnetization of CsEuFe$_4$As$_4$ at several representative temperatures [(a) $T$ = 10, 20, 30 and 40 K; (b) $T$ = 2 K]. The insets zoom in the interested regions that point to superconductivity and/or ferromagnetism.}
\end{figure}

Figure 4(a) shows the $M(H)$ curves at low temperatures. Above $T_{\mathrm{sc}}$, $M(H)$ is essentially linear, in accordance with paramagnetic state dominated by Curie-Weiss paramagnetism. At $T$ = 30 K, just below $T_{\mathrm{sc}}$, a typical superconducting loop appears, superimposing on the paramagnetic background. When the temperature is further decreased to 20 K, which is close to $T_{\mathrm{Curie}}$, the paramagnetic signals turn into a shape of Brillouin function, and the superconducting-like hysteresis loop is more obvious. Below $T_{\mathrm{Curie}}$, the $M(H)$ curves overall look like a ferromagnetic hysteresis loop, but it does not merge at higher magnetic fields (above the saturation field, $H_\mathrm{s}$). The latter phenomenon is clearly seen in the left inset of Fig. 4(b). This is exactly due to the existence of SC that gives rise to a flux pinning effect. Note that the saturation magnetization at 2 K is only 5.9 $\mu_\mathrm{B}$/Eu, somewhat smaller than the expected value of $gS$ = 7.0 $\mu_\mathrm{B}$/Eu for the full Eu$^{2+}$-spin ferromagnetic alignment. The lowered saturation magnetization is likely due to the superconducting screening effect.

Figure 5 plots the magnetic susceptibility $M/H$ as a function of temperature under $H$ = 1 kOe. Fitting the data from 50 to 300 K using an extended Curie-Weiss formula, $\chi$ = $\chi_0 + C$/($T-\theta$), yields three parameters, $\chi_0$ = 0.00142 emu mol$^{-1}$, $C$ = 7.27 emu K mol$^{-1}$ and $\theta$ = 23.1 K. From the $C$ value, one obtains the effective moment, $\mu_{\mathrm{eff}}$ = 7.63 $\mu_\mathrm{B}$/fu (fu denotes formula unit), which is close to the expected value of $\mu_{\mathrm{eff}} = g\sqrt{S(S+1)} \mu_\mathrm{\mathrm{B}}$ = 7.94 $\mu_\mathrm{\mathrm{B}}$/f.u. for Eu$^{2+}$ spins with $S$ = 7/2. The $\theta$ value, which represents a mean-field phase transition temperature, is significantly higher than the real transition temperature $T_{\mathrm{Curie}}$. This suggests that the interlayer magnetic interaction is much weaker, consistent with the dominant two-dimensional ferromagnetic interaction. The $\chi_0$ value should be mostly contributed from Pauli paramagnetism (since other sources of $T$-independent magnetism give a much lower value), therefore, one may estimate a density of state at Fermi level, $N(E_\mathrm{F})$ = 44 eV$^{-1}$ fu$^{-1}$, for CsEuFe$_4$As$_4$.

\begin{figure}
\includegraphics[width=8cm]{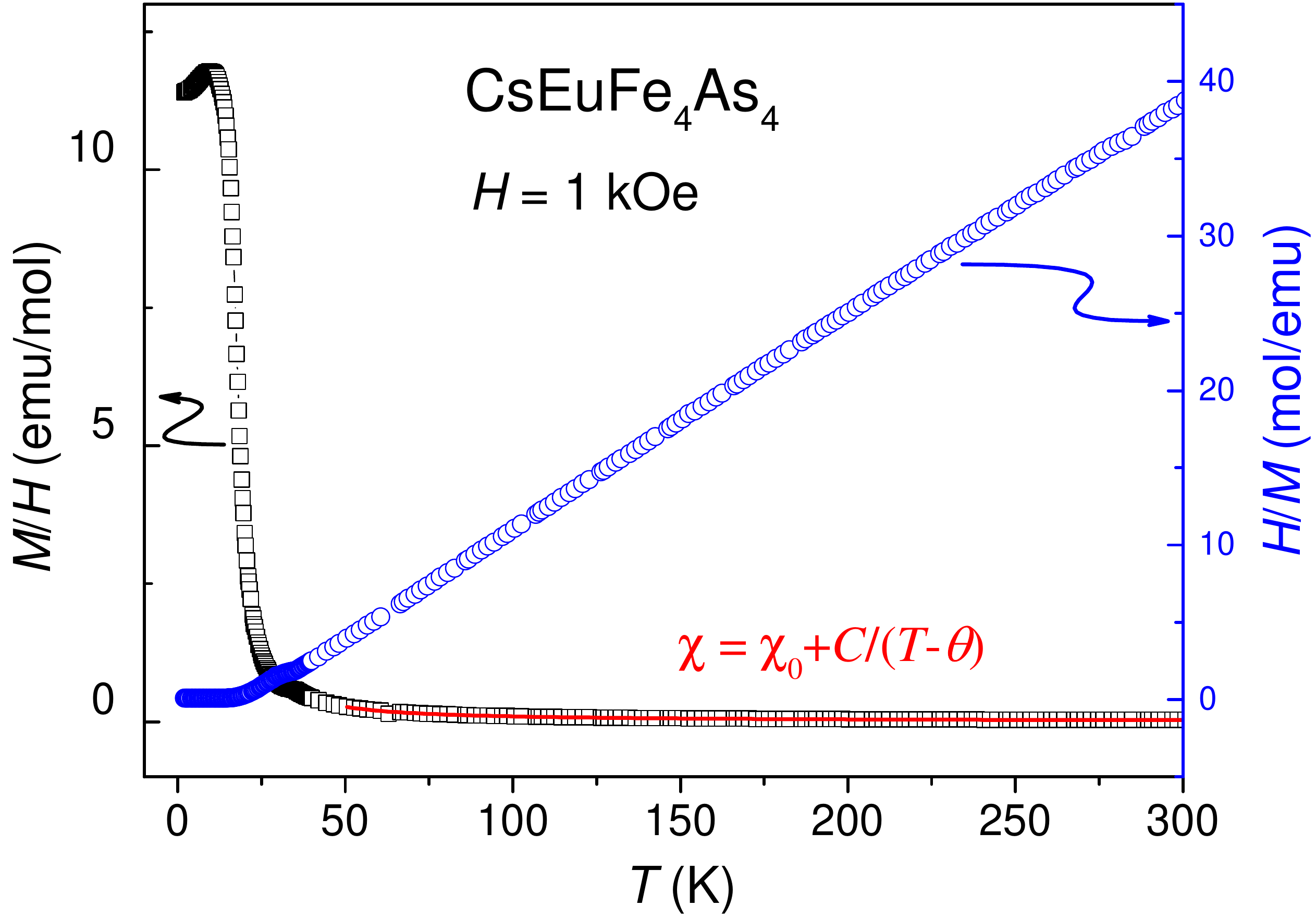}
\label{fig5} \caption{Magnetic susceptibility ($M/H$) of CsEuFe$_4$As$_4$ measured under $H$ = 1.0 kOe. The reciprocal of susceptibility is shown at the right axis.}
\end{figure}

\subsection{Specific heat}

Figure 6 shows the heat-capacity measurement result for CsEuFe$_4$As$_4$. Two anomalies at 15.2 and 35.2 K can be clearly seen, corresponding to the ferromagnetic and superconducting transitions, respectively. The obvious specific-heat jump further confirms the bulk nature of SC. As shown in the zoom-in plot, the $\Delta C$ value is as large as 6.5 J K$^{-1}$ mol$^{-1}$. Assuming the BCS weak-coupling scenario with $\Delta C$/($\gamma T_{\mathrm{sc}}$) = 1.43, the electronic specific-heat coefficient $\gamma$ is then estimated to be $\sim$130 mJ K$^{-2}$ mol-fu$^{-1}$, corresponding to $N(E_\mathrm{F})\approx$ 55 eV$^{-1}$ fu$^{-1}$ that basically agrees with the result from the magnetic measurement above. This large $\gamma$ value naturally explains the extraordinarily high specific heat beyond the Dulong-Petit limit at high temperatures, since the electronic specific heat contributes 39 J K$^{-1}$ mol$^{-1}$ at 300 K. As for the ferromagnetic transition, the expected jump for a second-order phase transition is absent at $T_{\mathrm{Curie}}$. However, the derivative of $C(T)$ (not shown here) shows a clear jump at $T_{\mathrm{Curie}}$, similar to the phenomenon previously observed in RbEuFe$_4$As$_4$\cite{Rb1144}. Therefore, the magnetic transition again represents a rare third-order phase transition, according to the earlier Ehrenfest classfication\cite{ehrenfest}. The upturn in $C/T$ [see Fig. 6(b)] below 20 K means that the magnetic ordering starts to ``nucleate" above $T_{\mathrm{Curie}}$, suggesting a strong two-dimensional character for the FM.

\begin{figure}
\includegraphics[width=7.5cm]{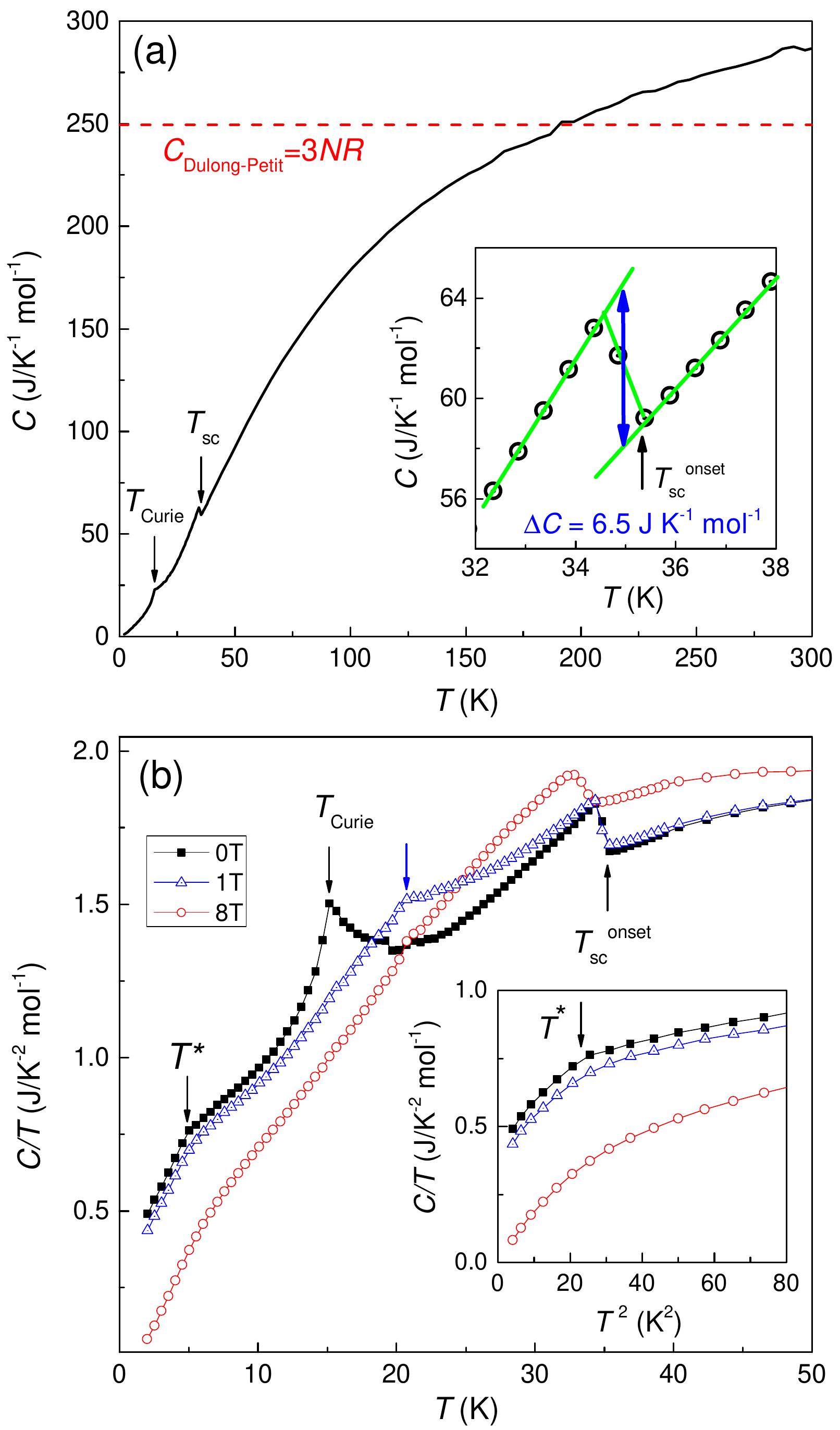}
\label{fig6} \caption{Heat capacity measurements for CsEuFe$_4$As$_4$. $T_{\mathrm{sc}}$, $T_{\mathrm{Curie}}$ and $T^*$ denote the superconducting, ferromagnetic and an unknown transitions, respectively. (a) Raw data of the specific heat, $C(T)$, at zero field. The inset zooms in the specific-heat jump at the superconducting transition. Panel (b) plots $C/T$ vs $T$ under 0, 1 and 8 T. The inset is a plot of $C/T$ vs $T^2$ at low temperatures.}
\end{figure}

Figure 6(b) plot the $C/T$ data under external magnetic fields. Upon applying magnetic fields, $T_{\mathrm{sc}}$ decreases very mildly, consistent with the magnetoresistance measurement above. However, $T_{\mathrm{Curie}}$ shifts to higher temperatures, further pointing to a ferromagnetic transition. The large zero-field $C/T$ values at low temperatures mainly come from the ferromagnetic magnon excitations, because they are greatly suppressed by the external fields. The residual $C/T$ value at $T \rightarrow$ 0 K tends to be zero (or a small value) even at 8 T [see the inset of Fig. 6(b)], suggesting a fully gapped SC. On the other hand, the high-temperature (above $T_{\mathrm{Curie}}$) $C/T$ values are significantly enhanced, which is associated with a field-induced magnetic ordering. An additional anomaly can be seen at $T^*$ = 5 K where $C/T$ shows a kink, coincident with the bifurcation of FCC and FCH curves in the magnetic measurement above.

\section{Discussions}
The above results demonstrate that CsEuFe$_4$As$_4$ is another new FMSC. Table~\ref{table2} compares the structural parameters and physical properties of the two sister compounds. While they share many similarities, there are still minor differences. Apart from the difference in the local structure of Fe$_2$As$_2$ layers described above, one notes that the $T_{\mathrm{sc}}$ and $T_{\mathrm{Curie}}$ values are negatively correlated. For CsEuFe$_4$As$_4$, the $T_{\mathrm{sc}}$ value is 1.3 K lower, but its $T_{\mathrm{Curie}}$ value is 0.5 K higher. Such a slight difference is likely to be related to the obviously short interatomic Eu$-$Fe distance in CsEuFe$_4$As$_4$, owing to the antagonism between SC and FM. Nonetheless, the FM in CsEuFe$_4$As$_4$ hardly suppresses $T_{\mathrm{sc}}$, because the $T_{\mathrm{sc}}$ value is reasonably between $T_{\mathrm{sc}}$ = 31.6 K for CsCaFe$_4$As$_4$ and $T_{\mathrm{sc}}$ = 36.8 K for CsSrFe$_4$As$_4$\cite{1144}.

\begin{table}
\caption{Comparison of structural and physical parameters of the two 1144-type ferromagnetic superconductors, CsEuFe$_4$As$_4$ (present work) and RbEuFe$_4$As$_4$\cite{Rb1144}. As1 and As2 denote the As atoms at Cs and Eu sides, respectively. $h_{\mathrm{As1}}$ ($h_{\mathrm{As2}}$) is the As1 (As2) height from Fe planes. $\alpha_{\mathrm{As1-Fe-As1}}$ ($\alpha_{\mathrm{As2-Fe-As2}}$) refers to As1$-$Fe$-$As1 (As2$-$Fe$-$As2) bond angle. $d_{\mathrm{Eu-Fe}}$ stands for the Eu$-$Fe interatomic distance.}
\label{table2}       
\begin{tabular}{cccccc}
\hline\noalign{\smallskip}
 Parameters & CsEuFe$_4$As$_4$& RbEuFe$_4$As$_4$\\
\noalign{\smallskip}\hline\noalign{\smallskip}
Space group     & $P4/mmm$& $P4/mmm$\\
Lattice parameter $a$ (\r{A}) & 3.9002(1) & 3.8896(1)\\
Lattice parameter $c$ (\r{A}) & 13.6285(4)& 13.3109(4)\\
$h_{\mathrm{As1}}$ (\r{A}) & 1.360& 1.300\\
$h_{\mathrm{As2}}$ (\r{A})  & 1.354& 1.386\\
$\alpha_{\mathrm{As1-Fe-As1}}$ ($^\circ$)     & 110.2& 112.5\\
$\alpha_{\mathrm{As2-Fe-As2l}}$ ($^\circ$)        & 110.5& 109.0\\
$d_{\mathrm{Eu-Fe}}$ (\r{A})        & 3.613& 3.684\\
\noalign{\smallskip}\hline\noalign{\smallskip}
$T_{\mathrm{sc}}$ (K)         &35.2&36.5\\
$T_{\mathrm{Curie}}$ (K)                     &15.5&15.0\\
$T^*$ (K)                     &5&5\\
$\mu_0$($dH_{\mathrm{c2}}/dT$)$_{T_{\mathrm{sc}}}$ (T/K)    &$-$5.3&$-$5.6\\
$\Delta C$ (J K$^{-1}$ mol$^{-1}$)             &6.5&7.5\\
$\gamma$ (mJ K$^{-2}$ mol$^{-1}$)        &130&150\\
$R_{\mathrm{H}}$(300K) (10$^{-9}$ m$^3$ C$^{-1}$)  &0.6&0.3\\
\noalign{\smallskip}\hline
\end{tabular}
\end{table}

The unsuppressed SC by FM strongly suggests that the coupling between superconducting Cooper pairs and Eu spins is avoided via a certain mechanism. Our previous explanation\cite{cao2011} still holds here for the FM+SC phenomenon. On the one hand, the Eu4$f$-Fe3$d_{yz/zx}$ coupling is vanishingly small, thus SC (from the $d_{yz/zx}$ electrons) is hardly affected. On the other hand, the Eu4$f$-Fe3$d_{x^{2}-y^{2}}$ and Eu4$f$-Fe3$d_{z^{2}}$ couplings remain finite, which give rise to effective in-plane and interlayer Ruderman-Kittel-Kasuya-Yosida (RKKY) exchange interactions, respectively. Particularly, the interlayer exchange interaction ($J_{\perp}$) may be greatly reduced, and even changes the sign, because the interlayer Eu interatomic distance is almost doubled as compared with EuFe$_2$As$_2$. Consequently, the Eu-spin FM can be realized in the presence of SC.

Finally, let us discuss the possible compromising way of FM+SC. Earlier theoretical studies presented two possibilities. One is so-called Fulde-Ferrell-Larkin-Ovchinnikov (FFLO) state\cite{ff,lo}, characterized by non-zero momentum for Cooper pairs. The other is called spontaneous vortex state (SVS)\cite{varma,tachiki}, which seems natural for a type II superconductor. Indeed, the bifurcation of magnetic susceptibility data in FCC and FCH processes in the temperature range of $T^*<T<T_{\mathrm{Curie}}$ points to a SVS, because spontaneous vortices naturally result in a thermal hysteresis. As for the FFLO state, in general, it requires more strict conditions including Pauli-limited $H_{\mathrm{c2}}$ with a large Maki parameter, small Fermi energy, clean-limit SC, etc. Nevertheless, as we argued previously\cite{Rb1144}, an FFLO state is possibly realized in CsEuFe$_4$As$_4$ and RbEuFe$_4$As$_4$ because the severe criteria may be satisfied. The anomaly at $T^*$ is possibly due to a transition from FFLO ($T<T^*$) to SVS ($T>T^*$). Further investigations with using the single-crystal samples are expected to supply more information on this interesting issue.

\section{Conclusions}
To summarize, we have synthesized a new Eu-containing 1144-type iron arsenide CsEuFe$_4$As$_4$, which is a sister compound of RbEuFe$_4$As$_4$\cite{Rb1144}. Although there are some differences in the crystal structure (for example, the Fe coordination in the Fe$_2$As$_2$ layers is nearly symmetric, in contrast with the obviously asymmetric Fe environment in RbEuFe$_4$As$_4$), CsEuFe$_4$As$_4$ shows similar physical properties with a superconducting transition at 35.2 K and an Eu-spin ferromagnetic ordering at 15.5 K. Our magnetic and specific-heat measurements unambiguously indicate a robust coexistence of bulk SC and strong FM, different from the FM+SC-like behaviors in doped EuFe$_2$As$_2$ systems. We believe that RbEuFe$_4$As$_4$ and CsEuFe$_4$As$_4$ are brand new FMSCs in which the interplay between SC and FM deserves further investigations.

\vspace{0.5cm}
\textbf{Acknowledgments}
This work was supported by National Science Foundation of China (11474252, 90922002 and 11190023) and National Basic Research Program of China (2012CB821404).


%

\end{document}